\documentclass{webofc}
\usepackage[varg]{txfonts}   % Web of Conferences font
\woctitle{International Nuclear Physics Conference 2013}
\begin{document}
\title{Strangeness Vector and Axial-Vector Form Factors \\ of the Nucleon}

\author{Stephen Pate\inst{1}\fnsep\thanks{\email{pate@nmsu.edu}} \and
        Dennis Trujillo\inst{1}}

\institute{Physics Department, New Mexico State University, Las Cruces NM 88003, USA}

\abstract{A revised global fit of electroweak $ep$ and $\nu p$ elastic scattering data has been 
performed, with the goal of determining the strange quark contribution to the
vector and axial-vector form factors of the nucleon in the momentum-transfer
range $0<Q^2<1$~GeV$^2$.  The two vector (electric and
magnetic) form factors $G_E^s(Q^2)$ and $G_M^s(Q^2)$ are strongly constrained by
$ep$ elastic scattering data, while the major source of information on the axial-vector
form factor $G_A^s(Q^2)$ is $\nu p$ scattering data.  Combining the two kinds of data
into a single global fit makes possible additional precision in the determination of
these form factors, and provides a unique way to determine the strange quark contribution to the
nucleon spin, $\Delta S$, independently of leptonic deep-inelastic scattering.  
The fit makes use of data from the BNL-E734, SAMPLE, HAPPEx, G0, and PVA4
experiments; we will also compare the result of the fit with recent data from MiniBooNE, and anticipate
how this fit can be improved when new data from MicroBooNE become available.}
\maketitle

\section{Overview}
The strange quark contribution to the elastic form factors of the nucleon has been the subject of intense experimental
scrutiny for several decades.  Experiments involving elastic scattering of neutrinos and anti-neutrinos 
(BNL E734~\cite{Ahrens:1986xe}),
and electrons (SAMPLE~\cite{Beise:2004py}, 
HAPPEx~\cite{Aniol:2004hp,Aniol:2005zg,Aniol:2005zf,Acha:2006my,Ahmed:2011vp}, 
G0~\cite{Armstrong:2005hs,Androic:2009zu}, 
and PVA4~\cite{Maas:2004ta,Maas:2004dh,Baunack:2009gy}) from nucleons and nuclei have explored
the strange quark presence in the nucleon by exploiting the electromagnetic and weak interactions in a variety of ways.
These special data permit a simultaneous determination~\cite{Pate:2003rk}
of the strange quark contribution to the electric ($G_E^s(Q^2)$), magnetic
($G_M^s(Q^2)$), and axial ($G_A^s(Q^2)$) form factors of the nucleon, which respectively allow us to understand how the
strange quark contributes to the distribution of charge, current, and spin inside the nucleon.  A global fit of
these data has been performed, which sets strong limits on the size and $Q^2$-dependence of $G_E^s(Q^2)$
and $G_M^s(Q^2)$, and points out the need for additional neutrino scattering data at low $Q^2$ to complete a measurement
of $G_A^s(Q^2)$.

\section{Strangeness Form Factors}
Since the strange and anti-strange quarks have opposite electric charges, then they contribute to the strangeness
electric form factor $G_E^s$ with opposite sign; therefore, if the $s$ and $\bar{s}$ distributions in the nucleon 
are the same, then $G_E^s$ will be zero.  Likewise, $s$ and $\bar{s}$ contribute oppositely to the strangeness
magnetic form factor $G_M^s$, and so any similarity between the $s$ and $\bar{s}$ distributions will drive $G_M^s$ to
small values.  However, $s$ and $\bar{s}$ have the same axial coupling, so if the $s$ and $\bar{s}$ distributions
are similar then a non-zero strangeness axial form factor $G_A^s$ can arise.

\section{Experiments}
To access the strangeness vector form factors of the nucleon, there has been a series of measurements of parity-violating
asymmetries in elastic electron-nucleon scattering~\cite{Armstrong:2012bi}, 
in which the electron is longitudinally polarized and the 
asymmetry in the cross section arises from the parity-violating nature of $Z$-exchange.  In forward scattering, the
PV asymmetries are most sensitive to the electric form factor, whereas in backward scattering there is a significant
contribution also from the magnetic and axial form factors.  In neutrino-nucleon scattering, on the other hand, the
most significant contribution comes from the axial form factor.  To date, the only useful measurement of the neutrino-proton
elastic scattering cross section is from BNL Experiment 734~\cite{Ahrens:1986xe}.  
Combining these two kinds of experiments together,
allows the simultaneous determination of the strange quark contribution to the vector and axial form factors~\cite{Pate:2003rk}.

\section{Determination of the Strangeness Form Factors at Specific Values of $Q^2$}
\begin{figure}[ht]
\begin{minipage}{16pc}
\includegraphics[trim = 0mm 50mm 0mm 0mm, clip, width=16pc]{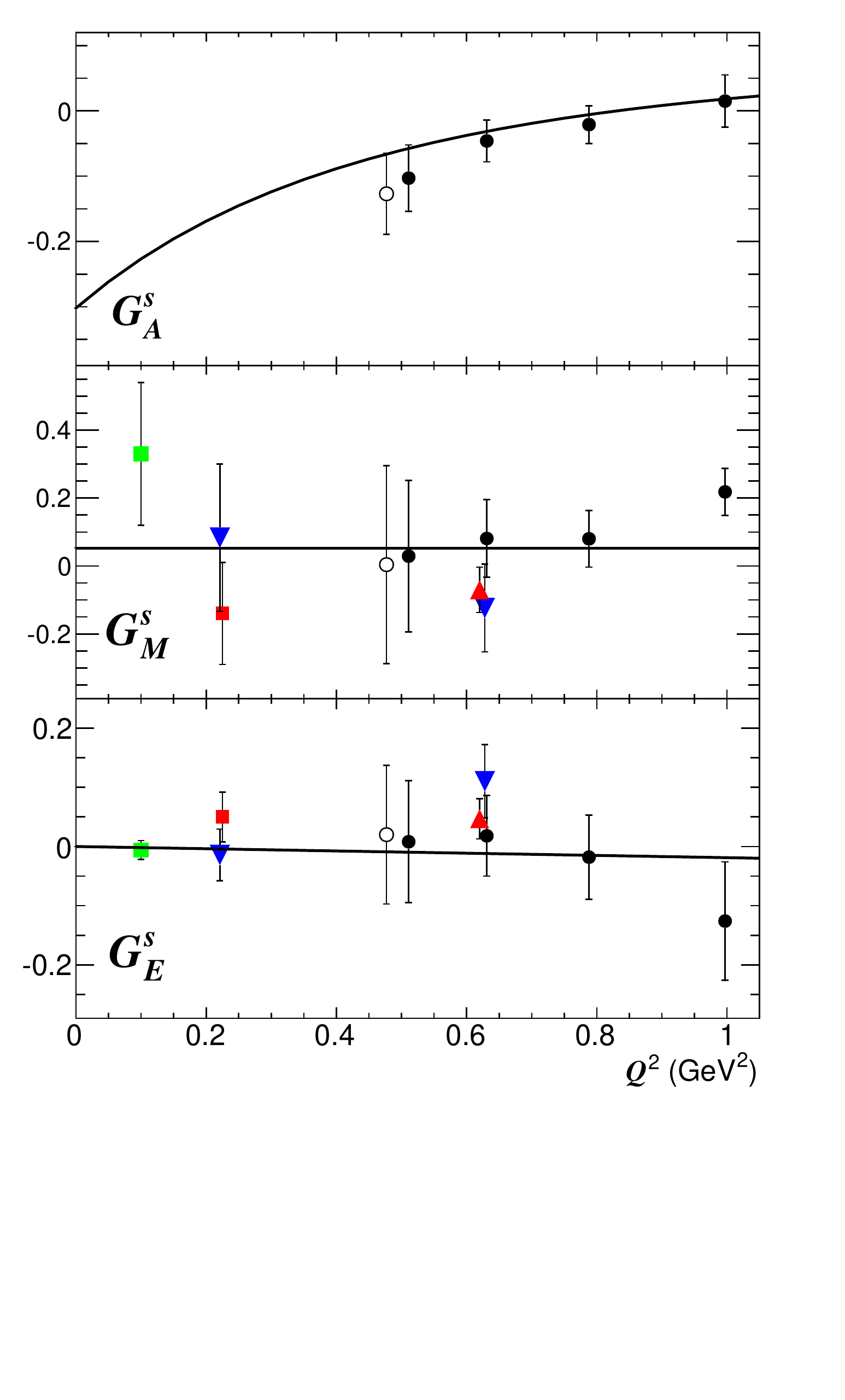}

\end{minipage}\hspace{2pc}%
\begin{minipage}{16pc}
\includegraphics[trim = 0mm 50mm 0mm 0mm, clip, width=16pc]{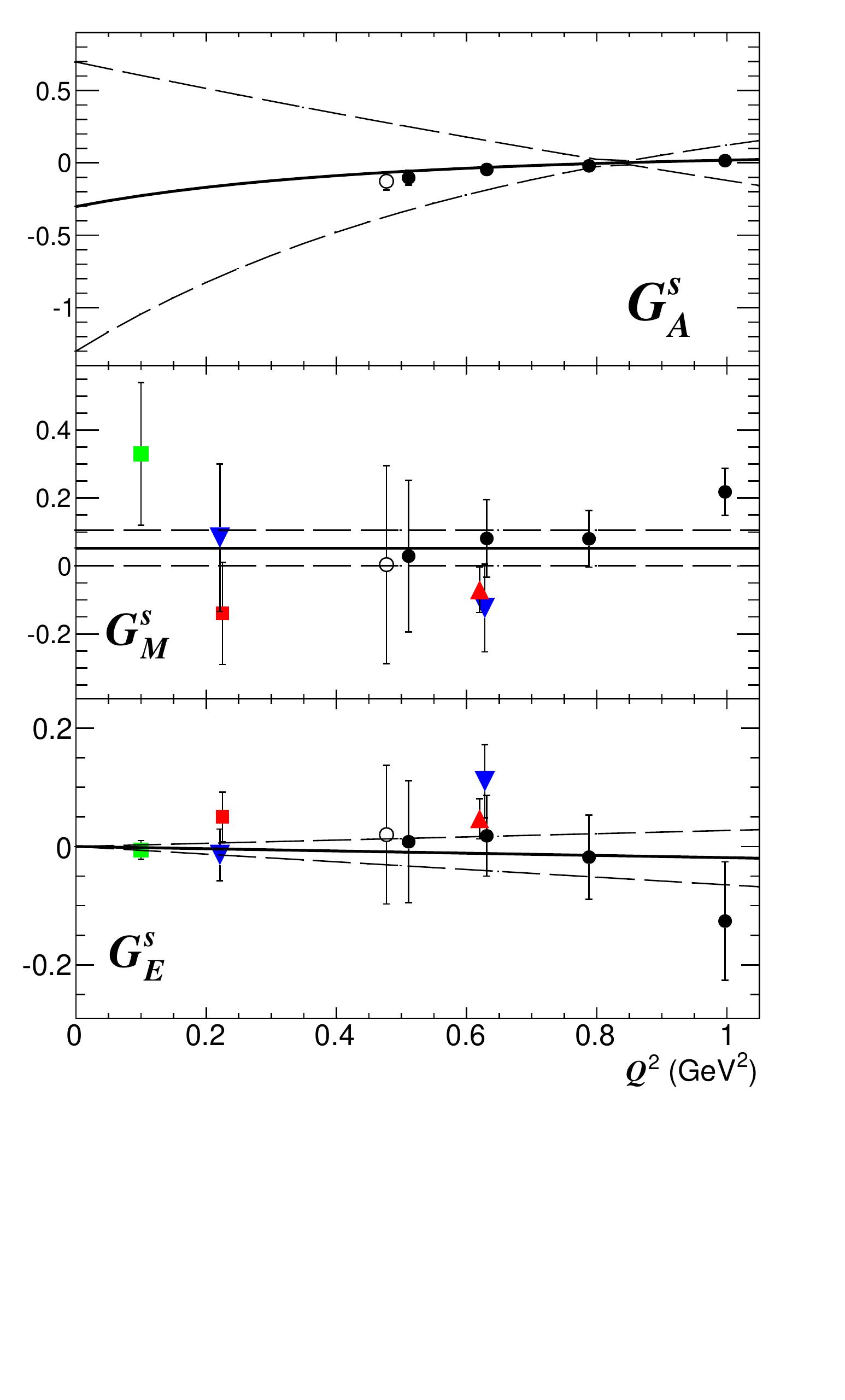}

\end{minipage} 
\caption{\label{fitfig} Results of the determination of $G_E^s$, $G_M^s$, and $G_A^s$ at
individual values of $Q^2$, and also from our global fit.  The separate determinations were done
by Liu et al.~\cite{Liu:2007yi} (green squares at 0.1 GeV$^2$), 
Androi\'{c} et al.~\cite{Androic:2009zu} (blue inverted triangles),
Baunack et al.~\cite{Baunack:2009gy} (red squares at 0.23 GeV$^2$), 
Ahmed et al.~\cite{Ahmed:2011vp} (red triangles at 0.62 GeV$^2$), and
Pate et al.~\cite{Pate:2008va} (open and closed circles).  The preliminary results of our global fit (see text)
are shown by the solid line; the 70\% confidence level limit curves for the fit are shown
as the dashed line in the right-hand panel.  The vertical scale for $G_A^s$ in the right-hand panel
has been adjusted to accommodate the limit curves of the fit.}
\end{figure}

Quite a number of analyses have been done, in which only the PV electron-scattering data are used and so only
the strangeness vector form factors $G_E^s$ and $G_M^s$ are extracted.  Using the techniques developed in
\cite{Pate:2008va}, it was possible to determine also the strangeness axial form factor at a number of points
where the HAPPEx and G0 experiments overlapped with the E734 data.  All of these determinations are shown in
Figure~\ref{fitfig}.  It is seen that both $G_E^s$ and $G_M^s$ are consistent with zero across the
range $0 < Q^2 < 1$~GeV$^2$, whereas in the axial form factor $G_A^s$ there is a signal for a non-zero
value as we approach $Q^2=0$.

\section{Global Fit of the Strangeness Form Factors}
Taken together, there are 48 data points from E734, SAMPLE, HAPPEx, G0 and PVA4.  Not all of these
are used in the determinations shown in FIgure~\ref{fitfig}.  To make use of all this data,
it necessary to assume functional forms for the strangeness form factors and perform a global fit.

Based on the results of the determinations at specific values of $Q^2$, we have
fit the form factors $G_E^s$, $G_M^s$, and $G_A^s$ in the range $0 < Q^2 < 1$~GeV$^2$ with this 
simple set of functional forms:
$$G_E^s = \rho_s\tau ~~~~~~~ G_M^s = \mu_s ~~~~~~~ G_A^s = \frac{\Delta S + S_A Q^2}{(1+Q^2/\Lambda_A^2)^2}$$
where $\tau = Q^2/4M_N^2$, $\rho_s \equiv (dG_E^s/d\tau)|_{\tau=0}$ is the strangeness radius, $\mu_s$ is
the strangeness magnetic moment, $\Delta S$ is the strange quark contribution to the nucleon spin,
and $\Lambda_A$ and $S_A$ determine the $Q^2$ dependence of $G_A^s$.

These are the simplest functions that are consistent with the determinations shown in Figure~\ref{fitfig}.
Since $G_E^s$ must be 0 at $Q^2$=0, then the lowest order term in $G_E^s$ must be linear in $Q^2$.
The lowest order term in $G_M^s$ is a constant, since $G_M^s$ need not be 0 at $Q^2=0$.  For
$G_A^s$, a more complex form is required.

The (preliminary) best values for the five fit parameters are given in Table~\ref{fit_table}.  We see that
the parameters representing the vector form factors ($\rho_s$ and $\mu_s$) are consistent with zero, as expected.
Strong limits are placed on the contribution of the strange quarks to the vector form factors throughout
this $Q^2$ range.  On the other hand,
$\Delta S$ is also consistent with 0 but the uncertainty is very large because there are no $\nu p$ or
$\bar{\nu} p$ elastic data at sufficiently low $Q^2$ to constrain it.  As a result the uncertainties in the
global fit to $G_A^s$ are very much larger than the uncertainties
in the separate determinations of $G_A^s$ in Figure~\ref{fitfig}.  We cannot determine $\Delta S$ in this method
until additional neutrino scattering data are obtained at low $Q^2$.

\begin{table}[ht]
\centering
\caption{Preliminary results for our 5-parameter fit to the 48 elastic neutrino- and PV electron-scattering data
points from BNL E734, HAPPEx, SAMPLE, G0, and PVA4.}
\label{fit_table}       % Give a unique label
\begin{tabular}{cc}
\hline
Parameter & Fit value \\ \hline
$\rho_s$ & $-0.071 \pm 0.096$ \\
$\mu_s$ & $0.053 \pm 0.029$ \\
$\Delta S$ & $-0.30 \pm 0.42$ \\
$\Lambda_A$ & $1.1 \pm 1.1$ \\
$S_A$ & $0.36 \pm 0.50$ \\ \hline
\end{tabular}
\end{table}

\section{Measurement of neutrino-proton elastic scattering at MicroBooNE}
MicroBooNE (\text{http://www-microboone.fnal.gov/}) is a new neutrino-scattering experiment under construction
at Fermilab, consisting of a 170-ton liquid-argon time projection chamber to be placed in the path of
a beam of approximately 1 GeV neutrinos.  This detector is ideal for observing neutrino-proton elastic scattering
events, as low-energy protons from these events can travel several centimeters in liquid argon; a measurement
down to $Q^2=0.08$ GeV$^2$ is possible.  A determination of $G_A^s$ down to such a low value of $Q^2$ would
permit a determination of $\Delta S$.  To estimate the level of uncertainty of such a measurement, a simulation
of $2\times 10^{20}$ protons-on-target was performed (about one running year), using reasonable event selection 
cuts.\footnote{Thanks to B. Fleming, J. Spitz, and V. Papavassiliou for providing this simulation.}  These simulated 
MicroBooNE cross section measurements were then fed back into our global fit program, and we observed the change in the
uncertainties in the fit parameters; see the table below.
\begin{table}[ht]
\centering
\caption{Improvement in uncertainties in fit parameters for $G_A^s$, when simulated MicroBooNE data are included
in the fit.}
\label{uboone_table}       % Give a unique label
\begin{tabular}{ccc}
\hline
Parameter & Existing Data & Including MicroBooNE \\ \hline
$\Delta S$ & $\pm 0.42$ & $\pm 0.038$ \\
$\Lambda_A$ & $\pm 1.1$ & $\pm 0.38$ \\
$S_A$ & $\pm 0.50$ & $\pm 0.071$\\ \hline
\end{tabular}
\end{table}
It is seen that a measurement of the strangeness axial form factor at MicroBooNE can have a dramatic effect
on this analysis, and we look forward to a determination of $\Delta S$ to come from this project in the next
few years.

\section{Acknowledgments}  This work was funded by the US Department of Energy, Office of Science.

%
% BibTeX or Biber users please use (the style is already called in the class, ensure that the "woc.bst" style is in your local directory)
\bibliography{INPC2013_Pate}

\begin{thebibliography}{16}

\bibitem{Ahrens:1986xe}
L.A. Ahrens et~al., Phys. Rev. \textbf{D35}, 785 (1987)

\bibitem{Beise:2004py}
E.J. Beise, M.L. Pitt, D.T. Spayde, Prog. Part. Nucl. Phys. \textbf{54}, 289
  (2005), \texttt{nucl-ex/0412054}

\bibitem{Aniol:2004hp}
K.A. Aniol et~al. (HAPPEx), Phys. Rev. \textbf{C69}, 065501 (2004),
  \texttt{nucl-ex/0402004}

\bibitem{Aniol:2005zg}
K.A. Aniol et~al. (HAPPEx), Phys. Lett. \textbf{B635}, 275 (2006),
  \texttt{nucl-ex/0506011}

\bibitem{Aniol:2005zf}
K.A. Aniol et~al. (HAPPEx), Phys. Rev. Lett. \textbf{96}, 022003 (2006),
  \texttt{nucl-ex/0506010}

\bibitem{Acha:2006my}
A.~Acha et~al. (HAPPEx), Phys. Rev. Lett. \textbf{98}, 032301 (2007),
  \texttt{nucl-ex/0609002}

\bibitem{Ahmed:2011vp}
Z.~Ahmed et~al. (HAPPEX collaboration), Phys.Rev.Lett. \textbf{108}, 102001
  (2012), \texttt{1107.0913}

\bibitem{Armstrong:2005hs}
D.S. Armstrong et~al. (G0), Phys. Rev. Lett. \textbf{95}, 092001 (2005),
  \texttt{nucl-ex/0506021}

\bibitem{Androic:2009zu}
D.~Androi\'{c} et~al. (G0), Phys. Rev. Lett. \textbf{104}, 012001 (2010),
  \texttt{0909.5107}

\bibitem{Maas:2004ta}
F.E. Maas et~al. (A4), Phys. Rev. Lett. \textbf{93}, 022002 (2004),
  \texttt{nucl-ex/0401019}

\bibitem{Maas:2004dh}
F.E. Maas et~al. (A4), Phys. Rev. Lett. \textbf{94}, 152001 (2005),
  \texttt{nucl-ex/0412030}

\bibitem{Baunack:2009gy}
S.~Baunack et~al. (A4), Phys. Rev. Lett. \textbf{102}, 151803 (2009),
  \texttt{0903.2733}

\bibitem{Pate:2003rk}
S.F. Pate, Phys. Rev. Lett. \textbf{92}, 082002 (2004), \texttt{hep-ex/0310052}

\bibitem{Armstrong:2012bi}
D.~Armstrong, R.~McKeown, Ann.Rev.Nucl.Part.Sci. \textbf{62}, 337 (2012),
  \texttt{1207.5238}

\bibitem{Liu:2007yi}
J.~Liu, R.D. McKeown, M.J. Ramsey-Musolf, Phys. Rev. \textbf{C76}, 025202
  (2007), \texttt{0706.0226}

\bibitem{Pate:2008va}
S.F. Pate, D.W. McKee, V.~Papavassiliou, Phys. Rev. \textbf{C78}, 015207
  (2008), \texttt{0805.2889}

\end{thebibliography}

\end{document}